\documentclass[11pt,english]{article}
\pdfoutput=1
\usepackage{graphicx,array}
\usepackage{cite}
\usepackage{color}
\usepackage{latexsym}
\usepackage{amsthm}
\usepackage{amsmath}
\usepackage[titletoc]{appendix}
\usepackage{enumitem}
\usepackage{amssymb}
\usepackage[unicode=true,
 bookmarks=true,bookmarksnumbered=false,bookmarksopen=false,
 breaklinks=false,pdfborder={0 0 1},backref=false,colorlinks=true]
 {hyperref}
\hypersetup{pdftitle={How Low Can Vacuum Energy Go When Your Fields Are Finite-Dimensional?},
 pdfauthor={Cao, Chatwin-Davies and Singh},
 citecolor=blue,linkcolor=blue,urlcolor=blue,citecolor=blue,linkcolor=blue}
\usepackage{breakurl}
\usepackage[hang,flushmargin]{footmisc} 

\def\simgt{\mathrel{\lower2.5pt\vbox{\lineskip=0pt\baselineskip=0pt
           \hbox{$>$}\hbox{$\sim$}}}}
\def\simlt{\mathrel{\lower2.5pt\vbox{\lineskip=0pt\baselineskip=0pt
           \hbox{$<$}\hbox{$\sim$}}}}

\newcommand{\be}{\begin{equation}}
\newcommand{\ee}{\end{equation}}
\newcommand{\bea}{\begin{eqnarray}}
\newcommand{\eea}{\end{eqnarray}}
\newcommand{\Eq}[1]{Eq.~(\ref{#1})}

\newcommand{\Fig}[1]{Fig.~\ref{#1}}

\newcommand{\mPl}{M_{\rm Pl}}
\newcommand{\lPl}{\ell_{\rm Pl}}

\definecolor{nicered}{rgb}{0.7,0.1,0.1}
\definecolor{nicegreen}{rgb}{0.1,0.5,0.1}
\hypersetup{colorlinks,citecolor=black,linkcolor=black,urlcolor=black}

\newcommand{\mrm}[1]{\mathrm{ #1 }}
\definecolor{purple}{rgb}{0.5,0,0.5}

\begin{document}
\baselineskip=12pt
\hfill CALT-TH-2019-017
\hfill

\vspace{2cm}
\thispagestyle{empty}
\begin{center}
{\LARGE\bf
How Low Can Vacuum Energy Go\\
When Your Fields Are Finite-Dimensional?
}\\
\bigskip\vspace{1cm}{
{\large ChunJun Cao, Aidan Chatwin-Davies, and Ashmeet Singh}\footnote{e-mail: \url{ccj991@gmail.com, aidan.chatwindavies@kuleuven.be, ashmeet@caltech.edu}\\Corresponding Author: A.S.}
} \\[7mm]
 {\it Joint Center for Quantum Information and Computer Science\\
University of Maryland, College Park, MD 20742, USA\\~\\
 
 KU Leuven, Institute for Theoretical Physics\\
 Celestijnenlaan 200D B-3001 Leuven, Belgium\\~\\
 Walter Burke Institute for Theoretical Physics\\
    California Institute of Technology \\
    1200 E. California Blvd., Pasadena, CA 91125} \\
 \end{center}
\bigskip
\centerline{\large\bf Abstract}

\begin{quote} \small
According to the holographic bound, there is only a finite density of degrees of freedom in space when gravity is taken into account.
Conventional quantum field theory does not conform to this bound, since in this framework, infinitely many degrees of freedom may be localized to any given region of space.
In this essay, we explore the viewpoint that quantum field theory may emerge from an underlying theory that is locally finite-dimensional, and we construct a locally finite-dimensional version of a Klein-Gordon scalar field using generalized Clifford algebras. 
Demanding that the finite-dimensional field operators obey a suitable version of the canonical commutation relations makes this construction essentially unique.
We then find that enforcing local finite dimensionality in a holographically consistent way leads to a huge suppression of the quantum contribution to vacuum energy, to the point that the theoretical prediction becomes plausibly consistent with observations.

\end{quote}

\vspace{2cm}

\begin{quote} \footnotesize
Essay written for the Gravity Research Foundation 2019 Awards for Essays on Gravitation.

Awarded Honorable Mention
\end{quote}

\newpage
\baselineskip=16pt
	
\setcounter{footnote}{0}

\section{Introduction: Gravity and Quantum Field Theory}
A quantum field theory has infinitely many degrees of freedom in any given region of space.
In the presence of gravity, when we try to excite such degrees of freedom that are supported on a compact region, many of the resulting states would collapse the region into a black hole.
Recall that a black hole has a finite amount of entropy which scales as the area of its horizon.
Therefore, any attempts to increase the region's entropy by creating further excitations would only increase the size of the resulting black hole, and hence also the size of its supporting region.
Such considerations suggest that the amount of entropy that can be localized in a compact region of space is finite \cite{Banks:2000fe,Bao:2017rnv, Banks2000, Fischler2000, Witten:2001kn,Dyson2002,Parikh:2004wh,Carroll:2018rhc}.
This idea is succinctly expressed through the holographic bound \cite{'tHooft:1993gx,Susskind:1994vu}, which says that the amount of entropy in a spacelike region $\mathcal{R}$ is bounded by the area of its boundary in Planck units,
\begin{equation}
S(\mathcal{R}) \leq \frac{|\partial \mathcal{R}|}{4 \lPl^2} .
\label{eqn:hbound}
\end{equation}

If we turn it around, the holographic bound, as well as its covariant generalizations \cite{Bousso:1999xy}, says that only finitely many degrees of freedom could have been localized to a compact region of space in the first place.
The von Neumann entropy of a maximally mixed state in a Hilbert space of dimension $\exp(|\partial \mathcal{R}|/4 \lPl^2)$ is enough to saturate the holographic bound, and so the degrees of freedom localized to $\mathcal{R}$ can have at most this number of orthogonal microstates.
In other words, the Hilbert space of a gravitating system is locally finite-dimensional.

At a first glance, it would therefore seem that quantum field theory is in conflict with the local finite dimensionality implied by gravity.
One way of addressing this conflict is to work within the framework of quantum field theory and introduce regulators so that it has effective finite dimension, e.g., through suitable infrared (IR) and ultraviolet (UV) cutoffs.
Another approach is to view the effective low energy behaviour of quantum field theory as something that must emerge from a theory that is intrinsically locally finite-dimensional.

We will explore the latter stance in this essay and construct a finite-dimensional version of an effective scalar field theory, in particular for which each mode cannot carry arbitrarily many excitations.
We will see that an automatic consequence of intrinsic finite dimensionality and the holographic bound is a tremendous suppression of the quantum contribution to vacuum energy compared to the prediction of conventional field theory.
Many authors before us have argued for observable consequences of holography in gravity, including corrections to vacuum energy \cite{Cohen:1998zx, Banks:1995uh, Horava:1997dd,Horava:2000tb}; this essay offers a fresh perspective on this line of reasoning through intrinsic finite dimensionality.

\section{Finite-Dimensional Effective Field Theory}
\subsection{Finite-Dimensional Field Operators}

In a conventional infinite-dimensional setting, such as the non-relativistic quantum mechanics of a single particle, classical conjugate variables $\phi$ and $\pi$ are promoted to linear Hilbert space operators which obey the Heisenberg canonical commutation relation (CCR)
\begin{equation}
  \label{CCR}
  [ \hat{\phi} , \hat{\pi} ] = i,
\end{equation}
where we have set $\hbar = 1$.
In a quantum field theory, the field and its conjugate momentum are operator-valued functions on spacetime which obey a continuous version of the CCR, labelled by spacetime points.

The Stone-von~Neumann theorem guarantees that there is an irreducible representation of \Eq{CCR}, which is unique up to unitary equivalence, on any infinite-dimensional Hilbert space that is separable (i.e., that possesses a countable dense subset) \cite{Kronz2005}.
However, in this case, the theorem also implies that the operators $\hat{\phi}$ and $\hat{\pi}$ must be unbounded.
There are therefore no irreducible representations of \Eq{CCR} on finite-dimensional Hilbert spaces. 

Instead, consider the following commutation relation due to Weyl \cite{weyl1950theory} on a finite-dimensional Hilbert space of dimension $d$:
\begin{equation}
\label{weylCCR}
e^{ - i \alpha \hat{\pi} } e^{i \beta \hat{\phi} } = e^{- i \alpha \beta } e^{i \beta \hat{\phi} } e^{- i \alpha \hat{\pi} }
\end{equation}
This is an exponentiated form of Heisenberg's CCR in the sense that, if the real parameters $\alpha$ and $\beta$ are chosen such that $\alpha\beta = 2\pi/d$, then \Eq{weylCCR} is equivalent to \Eq{CCR} in the limit as $d \rightarrow \infty$.
The operators $\hat \phi$ and $\hat \pi$ defined through Weyl's CCR \emph{do} admit an irreducible representation on a Hilbert space with finite dimension $d$.
Moreover, the representation is still unique up to unitary equivalence via the Stone-von~Neumann theorem, since a finite-dimensional Hilbert space is separable.

The generalized Clifford algebra (GCA) \cite{SanthanamTekumalla1976, Jagannathan1981, Jagannathan:1981ri,Jagannathan:2010sb, Singh:2018qzk} 
provides a simple way to write down the operators $\hat \phi$ and $\hat \pi$.
For example, let the dimension of Hilbert space be $d = 2l + 1$ for some non-negative integer $l$.
(The construction works when the dimension is even too, but we focus on odd values to streamline the notation.)
The GCA is generated by two unitary matrices
\begin{equation}
\hat A = \left(
\begin{array}{cccccc}
0 & 0 & 0 & \cdots & 0 & 1 \\
1 & 0 & 0 & \cdots & 0 & 0 \\
0 & 1 & 0 & \cdots & 0 & 0 \\
\vdots & \vdots & \vdots & \ddots & \vdots & \vdots \\
0 & 0 & 0 & \cdots & 0 & 0 \\
0 & 0 & 0 & \cdots & 1 & 0
\end{array} \right) \qquad \hat B = \left(
\begin{array}{cccc}
\omega^{-l} & 0 & \cdots & 0 \\
0 & \omega^{-l+1} & \cdots & 0 \\
\vdots & \vdots & \ddots & \vdots \\
0 & 0 & \cdots & \omega^l
\end{array} \right),
\end{equation}
which satisfy the commutation relation $\hat A \hat B = \omega^{-1} \hat B \hat A$ and multiplicative closure $\hat A ^{d} = \hat B ^{d} = \mathbb{I}_{d}$, where $\omega = \exp(2\pi i/d)$.
The identification $\hat A \equiv \exp(-i\alpha \hat\pi)$ and $\hat B \equiv \exp(i \beta \hat \phi)$ then realizes \Eq{weylCCR}.
These conjugate operators $\hat{\phi}$ and $\hat{\pi}$ from the GCA each have a bounded, linearly-spaced, discrete spectrum of dimensionless eigenvalues.
In the infinite-dimensional limit, they reduce to the usual conjugate operators with unbounded spectra and obey the Heisenberg CCR.

\subsection{Vacuum Energy}

Let us now use these finite-dimensional conjugate operators to construct a finite-dimensional version of a scalar field theory.
Consider first a scalar field in a three-dimensional box of side length $L$ with the usual Klein-Gordon Hamiltonian, which we can decompose in terms of its Fourier modes,
\begin{equation}
\hat{H} = \sum_{\vec k} \hat{H}_{\vec{k}} = \frac{1}{L} \sum_{\vec k}  \left( \frac{1}{2} \hat{\pi}^{2}_{\vec{k}} + \frac{1}{2}\Omega^{2}_{k} \hat{\phi}^{2}_{\vec{k}} \right).
\end{equation}
This Hamiltonian describes a number of decoupled quantum harmonic oscillators, one for each mode $\vec k$ with natural frequency $\Omega_k$.
We have pulled out a factor of $1/L$ in the equation above so that the $\hat \phi_{\vec k}$'s and $\hat \pi_{\vec k}$'s are dimensionless. 
Similarly, each $\Omega_k \equiv kL$ is a dimensionless frequency, where $k \equiv |\vec{k}|$.
Since we have cast the Hamiltonian in terms of dimensionless operators, we can obtain a finite-dimensional theory by simply replacing the $\hat \phi_{\vec k}$'s and $\hat \pi_{\vec k}$'s with the (dimensionless) finite-dimensional operators described above.
Letting the Hilbert space dimension $d_{k}$ of each mode go to infinity restores the original Klein-Gordon theory.
For finite dimension $d_k$, however, the spectrum of each $\hat{H}_{\vec k}$ is not linearly spaced and possesses both maximum and minimum eigenvalues which depend on the values of $\Omega_k$ as well as $d_k$ \cite{Singh:2018qzk}. 

To fix the $d_k$'s, we come back to the question of creating black holes.
A gravitationally-acquainted effective field theory should cut off below any excitations that would collapse into black holes.
Therefore, it is natural to impose that the largest energy eigenvalue for each mode $k$ should not exceed the Schwarzschild energy of the box.
The largest energy eigenvalue is fixed according to the finite-dimensional construction of conjugate variables introduced in the previous section.
As is discussed in Ref.~\cite{Singh:2018qzk}, the largest eigenvalue of each finite-dimensional conjugate operator, $\hat{\phi}_{\vec{k}}$ and $\hat{\pi}_{\vec{k}}$, is $l \sqrt{2 \pi/ (2l+1)}$.
Then, from Horn's inequalities \cite{horn1962,KnutsonTao}, one can consequently show that the largest eigenvalue of $\hat{H}_k$ is tightly bounded by
\begin{equation}
E_{\mrm{max}}(k)  \leq  \frac{\pi \Omega_{k}^{2} d_{k}}{4L}.
\end{equation}
By demanding that $E_\mrm{max}(k)$ be less that $\sim L \mPl^2/4$, we arrive at
\begin{equation}
d_{k} \lesssim \frac{L^{2} M^{2}_{pl}}{\pi \Omega^{2}_{k}}.
\label{eqn:dbound}
\end{equation}
In particular, this bound suppresses the dimension of high-frequency modes and defines a smallest mode residing at $k = \mPl/\sqrt{\pi}$, for which $d_k = 1$.
Note that this construction is inherently different from simply truncating every Klein-Gordon mode's usual spectrum, each of which is that of an infinite-dimensional harmonic oscillator.

\begin{figure}[h]
\includegraphics[width=\textwidth]{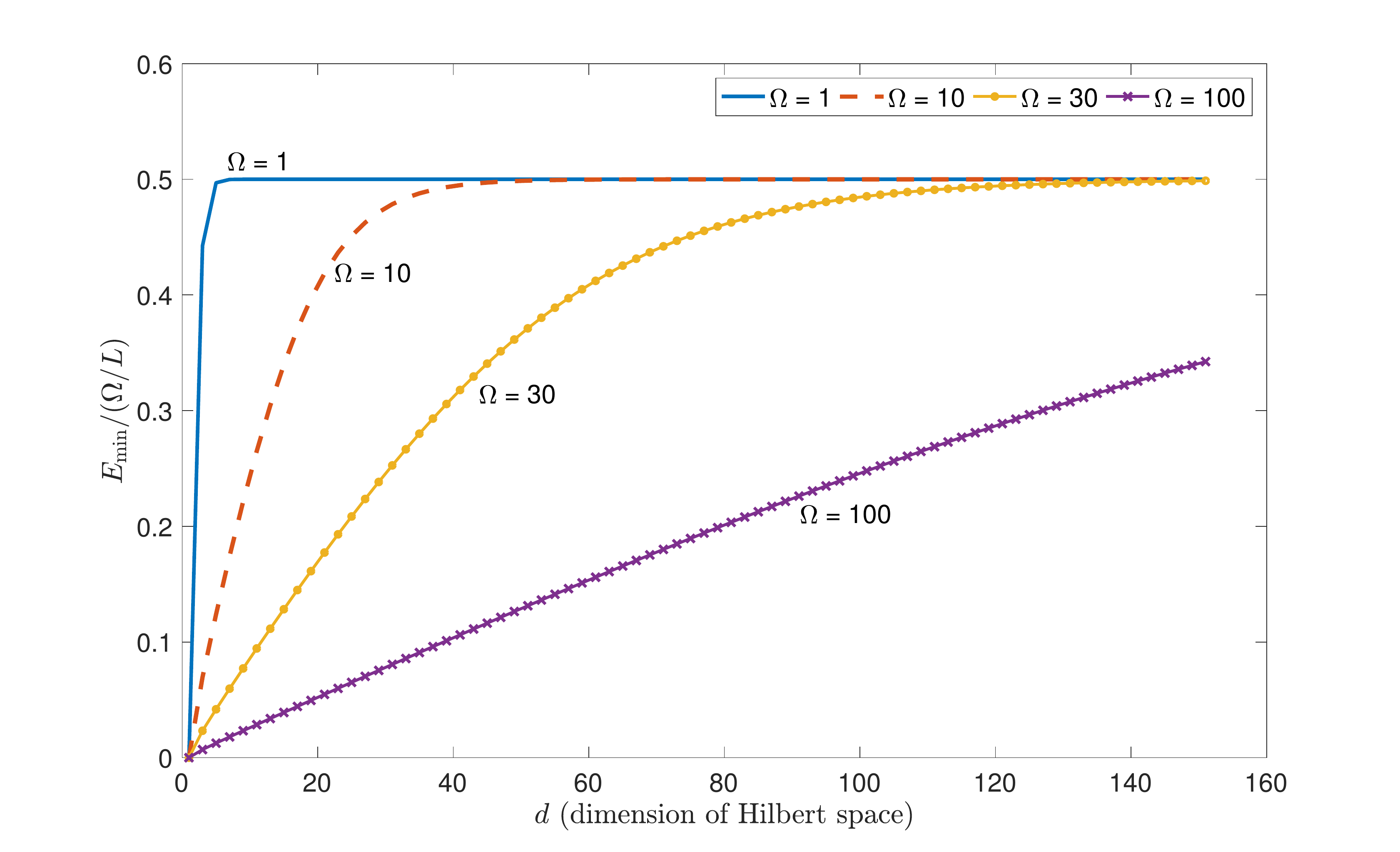}
\caption{Minimum energy eigenvalue for a finite-dimensional field mode, normalized by $\Omega/L$, as a function of the dimension $d$ and for several values of $\Omega$. }
\label{fig:oscillator_min}
\end{figure}

Let us now examine the minimum energy eigenvalue of each mode.
$E_{\mrm{min}}(k)$ is always bounded above by $k/2$, the zero point energy of an infinite dimensional oscillator, and lowering the value of $d_k$ lowers the value of $E_{\mrm{min}}(k)$; this is illustrated in \Fig{fig:oscillator_min}.
Therefore, the bound \eqref{eqn:dbound} also suppresses $E_{\mrm{min}}(k)$, with the suppression becoming increasingly severe at higher frequencies.

In summary, we find that imposing a finite dimension that prevents each mode from exceeding the box's Schwarzschild energy reduces the ground state energy of each field mode.
The quantum contribution to total vacuum energy density will consequently be lowered as well, even when summing over modes all the way to the Planck scale.
While our finite-dimensional field-in-a-box is not a precise cosmological model, we can get a sense for what the size of the effect might be for our Hubble patch by taking the box size $L$ to be the current Hubble radius.
The resulting vacuum energy density that we compute is
\begin{equation}
\begin{aligned}
\rho_0^{GCA} &= \frac{1}{L^3} \sum_{\vec{k}} E_{\mrm{min}}(k) \\
&\approx \frac{1}{L^3} \int_{L^{-1}}^{{\mPl}/\sqrt{\pi}} dk ~ 4 \pi k^{2}\left(\frac{L}{\pi}\right)^3  E_{\mrm{min}}(k) ~ \lesssim ~ (10^4 ~\mrm{GeV})^4 .
\end{aligned}
\end{equation}
This is 60 orders of magnitude lower than the na\"ive counting of vacuum energy density contribution from a free Klein-Gordon field,
\begin{equation}
\begin{aligned}
\rho_0^{KG} &= \frac{1}{L^3} \sum_{\vec{k}} \frac{k}{2} \\
&\approx \frac{1}{L^3} \int_{L^{-1}}^{{\mPl}} dk ~ 4 \pi k^{2} \left(\frac{L}{\pi}\right)^3 \frac{k}{2} ~ \sim ~ (10^{19}~\mrm{GeV})^4.
\end{aligned}
\end{equation}

\section{Conclusion: Hilbert Space and Holography}

While an intrinsically finite-dimensional version of a scalar field results in a vastly smaller vacuum energy compared to the original infinite-dimensional theory, it is still many orders of magnitude above the observed value, $\rho_0 \lesssim (10^{-11}~\mrm{GeV})^4$ \cite{Perlmutter:1998np,Riess:1998cb}.
However, this is because our simple estimate still counts many more states than allowed by the holographic bound.
This can be seen by setting $d_k$ equal to the bound \eqref{eqn:dbound} and computing the dimension, $D$, of the total Hilbert space:
\begin{equation}
\log D = \sum_{\vec{k}} \log d_k \approx \int_{L^{-1}}^{\mPl/\sqrt{\pi}} dk~4 \pi k^2 \left( \frac{L}{\pi} \right)^3 \log d_k \sim (L\mPl)^3 + \mathcal{O}(\log(L\mPl))
\end{equation}
According to the holographic bound, this should be no more than $S \sim (L \mPl)^2$.

Local finite dimension according to \Eq{eqn:dbound} alone is therefore not the end of the story.
There will also be a holographic depletion of states, which should be strongest in the UV \cite{Yurtsever:2003ii,Aste:2004ba}.
This can be understood heuristically by noting that the density of field theoretic degrees of freedom is observed to scale extensively in the IR, and also that many otherwise-valid states would collapse to form black holes in the UV.
For example, exciting every Klein-Gordon mode up to $k \sim 1~\mrm{meV}$ is already enough to reach the Schwarzschild energy of our universe-sized box.

In a crude attempt to model this holographic depletion of states, we can try modifying the density of modes $g(k)~dk = 4 \pi k^2~dk$ by setting
\begin{equation}
\tilde g(k) = \left\{
\begin{array}{ll}
4 \pi k^2 & L^{-1} < k < k_* \\
2\pi k & k_* < k < {\mPl}/\sqrt{\pi}
\end{array} \right. .
\end{equation}
Taking the crossover to lie at $k_* \sim 1~\mrm{meV}$, we find that the vacuum energy is reduced to $\rho_0^{GCA} \lesssim (10^{-10}~\mrm{GeV})^4$, which is consistent with the observed value of vacuum energy to within an order of magnitude.
This result should be taken with a grain of salt, however, due to the delicate interplay between the finite oscillator dimensions $d_k$ and the density of modes $\tilde g(k)$, which we have yet to investigate in detail.
Nevertheless, the calculation discussed here for vacuum energy illustrates that taking finite dimensionality and holography together seriously can have important predictive consequences for gravity.

\begin{center} 
 {\bf Acknowledgments}
 \end{center}

We would like to thank Sean Carroll and Grant Remmen for helpful discussions. C.C. acknowledges the support by the U.S. Department of Defense and NIST through the Hartree Postdoctoral Fellowship at QuICS. A.C.-D. is currently supported in part by the KU Leuven C1 grant ZKD1118 C16/16/005, the National Science Foundation of Belgium (FWO) grant G.001.12 Odysseus, and by the European Research Council grant no. ERC-2013-CoG 616732 HoloQosmos.
A.S. is funded in part by the Walter Burke Institute for Theoretical Physics at Caltech, by DOE grant DE-SC0011632, and by the Foundational Questions Institute.

\bibliographystyle{utphys}
\bibliography{cc_finite_7}

\end{document}